\documentclass[aps,pra,showpacs,twoside,twocolumn,10pt]{revtex4-1}
\usepackage[colorlinks=true, citecolor=red, urlcolor=blue ]{hyperref}
\usepackage{epsfig,newlfont,amssymb,amsfonts,amsmath,bm,subfigure,palatino,mathtools,amsthm,braket,times,soul,enumitem,color}
\usepackage[normalem]{ulem}

\usepackage{blindtext}
\usepackage{graphicx}
\usepackage{amsmath}
\usepackage{bm}
\usepackage{hyperref}
\usepackage{geometry}
\usepackage{amsthm}
\usepackage{gensymb}
\usepackage{physics}
\begin{document}

\title{Quantum entanglement and anthropology}
\author{Ujjwal Sen}
\affiliation{Harish-Chandra Research Institute, A CI of Homi Bhabha National Institute, Chhatnag Road, Jhunsi, Allahabad 211 019, India}
\date{01 April 2022}

\begin{abstract}
   We find that the set of local quantum operations and classical communication for multiparty quantum states can be considered as analogous to online meetings between members of a population. Moreover, monotonicity properties of quantum and classical correlations of quantum states of shared systems also carry over to relations between members of the population, giving rise to what may be termed as a second law of anthropology.
\end{abstract}
\maketitle

Entanglement~\cite{Horodecki;Sreetama-Titas} of shared quantum states is a useful resource in quantum technologies~\cite{Nielsen-Chuang;Preskill;Bruss-Leung}. It is often, maybe not always~\cite{4:04AM27Dec2021}, the diesel that runs quantum devices. 

Anthropology is  ``the science that deals with the origins, physical and cultural development, biological characteristics, and social customs and beliefs of humankind''~\cite{Utkal-gyalo-Ashoke}.

The distant laboratories paradigm is a very useful concept in the practice and theory of entanglement, and indeed in a large part of quantum information. When two or more parties share a quantum state, they are usually able to perform only local quantum operations in their own labs, and communicate classically between the labs. This is a severely restricted set of operations in comparison to what is possible, in principle, on the multiparty system. This is because transfer of quantum bits between the labs is disallowed. This restricted set of operations is called ``LOCC'' - local quantum operations and classical communication. 

Under this restricted set of operations, the ``entanglement content'' of a multiparty quantum system can only go down or remain unchanged, pointing to a form of the ``second law'', where ``entanglement content'' need to be carefully defined. In particular, if there are only two labs that are operating, we can conceptualize the entanglement content of the bipartite quantum states shared between the two labs by using the concept of entanglement cost~\cite{ma-tui-ayaman-kyano-hali} or distillable entanglement~\cite{patir-buke-charan-rekhe}, respectively being the asymptotic rates at which the corresponding state can be created from singlets or vice versa. Another useful measure of bipartite entanglement (that also works in the multiparty domain) is the relative entropy of entanglement~\cite{hridayer-gaan-shikhe-to-gaye-go-sobai}, defined as the relative entropy distance~\cite{lal-phnite-sada-moja} of the corresponding bipartite quantum state to the closed and convex set of unentangled quantum states. In case there are more than two labs, one can invoke multiparty entanglement measures (see e.g.~\cite{palatak-mon-shunya, palatak-mon-ek,palatak-mon-dui, palatak-mon-tin}). 

Within the distant laboratories paradigm, therefore, entangled states that have already been created beforehand may remain entangled, with its entanglement content either remaining the same or getting diminished. In the extreme case, the state can also  become separable (unentangled). However, an unentangled state can never become entangled in this set-up. This is the statement of the \emph{second law of shared quantum systems}, and can, somewhat loosely, be stated as\\

``entanglement does not grow under LOCC.''\\

Multiparty quantum states, whether they are entangled or otherwise, can certainly be classically correlated.
Defining classical correlations often leads one to slippery grounds, and one way to handle it is to consider the relative entropy distance of the closest separable state to the tensor product of the local densities~\cite{hridayer-gaan-shikhe-to-gaye-go-sobai}. If the state under consideration is separable, then it itself serves as the closest separable state. However, whatever is the definition of classical correlation content of a bipartite or multiparty quantum system, it can, in general, increase under LOCC. 

Consider now a population of \(n\) humans belonging to the set \(\mathcal{P} = \{h_i | i=1, \ldots, n\}\). Consider now a subset \(\mathcal{R}_\alpha\) of \(\mathcal{P}\), that contains at least two members of \(\mathcal{P}\) and in which, any two members are relatives to at least \(\alpha\) percent  of closeness, for a certain fixed value of \(\alpha\). The degree of closeness for a pair of members can be defined in the usual way.
E.g.,  Alice and her mother are relatives of 50\% closeness, while she and her grandmother are relatives of 25\% closeness. Alice and her partner, if existing and if they have met at least once offline,
are relatives of 100\% closeness, and in that case, she is a relative of 50\% closeness with her mother-in-law, etc. 

Consider another subset \(\mathcal{F}_\beta\) of \(\mathcal{P}\) that contains at least two members of \(\mathcal{P}\) and in which, any two members are friends to at least \(\beta\) percent of closeness. Two people are said to be friends of 100\% closeness, if they have met at least once, either in person or online.
For others, the percentages are lower. E.g., if Alice-Bob, Bob-Charu, and Charu-Debu are, for each pair, friends of 100\% closeness, and if no other pair have ever met even online, then Alice and Charu 
are friends of 50\% closeness, and Alice and Debu are friends of 25\% closeness. 

We can define the \(\alpha\) of a set of relatives \(\mathcal{R}_\alpha\) as the degree of closeness of the set \(\mathcal{R}_\alpha\). Similarly, we can define 
\(\beta\) of a set of friends \(\mathcal{F}_\beta\) as the degree of closeness of the set \(\mathcal{F}_\beta\).

We now see that the degree of closeness of a set, \(\mathcal{R}_\alpha\), of relatives is an invariant, under the restriction that the members of \(\mathcal{R}_\alpha\) are only allowed to meet online. On the other hand, the degree of closeness of a set, \(\mathcal{F}_\beta\), can increase or remain invariant under the same restriction. 
This can be considered as a \emph{second law of anthropology}, and can, somewhat carefully, be stated as\\

``relatives cannot be created online.''\\

Therefore, the degree of closeness of a set of relatives under the restriction of only online meetings behaves similarly as the entanglement content of a multiparty quantum state under the restriction of LOCC. Indeed, although entanglement does decrease under LOCC, the diminishing happens only if the dynamics involved is open.

On the other hand, the degree of closeness of a set of friends meeting  online or offline has a behavior that is similar to that of classical correlations of multiparty quantum states under LOCC.

\begin{acknowledgments}
We acknowledge discussions with several junior members of the quantum information and computation group at the Harish-Chandra Research Institute about \emph{different} aspects of ``LOCC''. We acknowledge partial support from the Department of Science and Technology, Government of India through the QuEST  grant (grant number DST/ICPS/QUST/Theme-3/2019/120).
\end{acknowledgments}

\end{document}